\documentclass[aps,twocolumn,showemail,preprintnumbers,floatfix]{revtex4}
\usepackage{amsmath}
\usepackage{graphics}

\newcommand \be{\begin{equation}}
\newcommand \ee{\end{equation}}

\begin{document}

\title{\textbf{The electron-phonon interaction from fundamental local gauge
symmetries in solids}}
\author{C.A. Dartora$^{1}$}
\email{cadartora@eletrica.ufpr.br}
\author{G.G. Cabrera$^{2}$}
\email[Corresponding author: ]{cabrera@ifi.unicamp.br}
\affiliation{$^{1}$ Electrical Engineering Department, Federal University of Parana
(UFPR), C.P. 19011 Curitiba, 81.531-970 PR, Brazil \\
$^{2}$ Instituto de F\'{\i}sica \ \ `Gleb Wataghin', Universidade Estadual
de Campinas (UNICAMP), Campinas 13.083-859 SP, Brazil\\
}

\begin{abstract}
The elastic properties of solids are described in close analogy with General
Relativity, by locally gauging the translational group of space-time.
Electron interactions with the crystal lattice are thus generated by
enforcing full gauge invariance, with the introduction of a gauge field.
Elementary excitations are associated with the local gauge, contrasting to
the usual interpretation as being Goldstone bosons emerging from global
symmetry breaking. In the linear limit of the theory, the gauge field
displays elastic waves, that we identify with acoustic phonons, when the
field is quantized. Coupling with the electronic part of the system yields
the standard electron-phonon interaction. If spin-orbit effects are
included, unusual couplings emerge between the strain field and the
electronic spin current, leading to novel physics that may be relevant for
spintronic applications.

KEYWORDS: non-linear elasticity; General Relativity; gauge symmetry;
non-abelian gauge fields; electron-phonon interaction; spin currents

\end{abstract}

\maketitle


\section{Introduction}

In the physics of condensed matter, the concept of collective excitations
gives rise to a large number of quasi-particles associated with the quanta
of these excitations. Some well known examples, among the most important,
are the vibrational modes of the atomic lattice, whose quanta are called
phonons, and spin waves whose quanta are magnons\cite{[kittel]}. There are
two prominent and, in many respects, inequivalent ways to get the physics of
collective excitations: i) by means of the so-called \emph{emergence}
principle, and ii) by invoking \emph{gauge symmetry}. For the former, the
detailed interactions between \ `truly' elementary particles are considered
at the very beginning, leading to the \ `emergence' of their collective
excitations and the quantum field describing it. For the latter, which is
well known to high energy and particle physicists, it is the gauge symmetry
principle which will tell us about the allowed interactions and what kinds
of particles can be considered as elementary\cite{[weinberg]}.

Historically, in condensed matter physics, the most relevant collective
interactions were firstly obtained via the emergence principle. But
important cooperative phenomena, such as superfluidity and
superconductivity, are now recognized as gauge theories, and the gauge
principle is being currently used with great success in the condensed matter
province\cite{reference}. This way, new visions of the field are opened,
including symmetry breaking and generation of mass via the Higgs mechanism,
concepts that show the unified character of physical laws. This fact leads
Weinberg to state literally that \emph{``a superconductor is simply a
material in which the electromagnetic gauge invariance is spontaneously
broken"}\cite{[weinberg]}.

In a different road, but following the same paradigm, spin interactions of
non-relativistic Pauli electrons were obtained by considering the full gauge
symmetry group of the Pauli-Schr\"{o}dinger theory\cite{[froehlich]}, with
no explicit reference to the Dirac equation. Within the same framework,
Dartora and Cabrera showed that the spin-current density operator can be
correctly defined using gauge invariance and Noether's theorem\cite%
{[dartora]}. The above quantity has fundamental relevance in the field of
spintronics \cite{[spin1],[spin2],[spin3]}, allowing to explain several
interesting phenomena that include the spin Hall effect and the
spin-transfer torque mechanism\cite{[hall],[hall1], [hall2]}. The authors
have also studied magnetic excitations (magnons)\ following a similar
approach, gauging the $SU(2)$ group\cite{magnons}.

Many open questions remain to be elucidated in this cross field between
condensed matter and high energy physics. In particular, the theory of
gravity has resisted a successful quantization, a problem which is believed
to be relevant at very high energy scale. However, some realistic and exotic
scenarios can be explored and tested in the low energy regime of condensed
matter. For instance, the study of electronic transport in graphene has
shown that, under certain conditions, a slowly moving electron behaves as a
relativistic Dirac electron in a space-time structure that mimics gravity
theory\cite{[mesaros]}. In another direction, artificial metric spaces have
been created in optical systems to simulate exotic space-time geometries
\cite{[metric1]}. Acoustic \ `black holes' display many of the properties
that are attributed to black holes of general relativity, as shown by Unruh%
\cite{unruh}. Topological defects in graphene have been described by gauge
fields which couple to the Dirac equation for Dirac electrons\cite{dirac}.
All the above facts suggest that some condensed matter systems can be
engineered with a prescribed set of gauge symmetries, allowing to test
several aspects of high energy physics and gravity theory, otherwise
impossible to do within current technological limitations.

In this manuscript, we follow the gauge paradigm, suggesting that
interactions in condensed matter systems are consequence of local gauge
invariance and follow from general symmetry principles. This assumption
serves as a guidance to study the motion of electrons in a solid with
elastic properties. Mutual interactions result from local symmetry
properties, with the gauge field describing the elementary excitations of
the solid. In fact, we proceed by requiring invariance under local
space-time translations of our condensed matter system, leading in a natural
way to the introduction of gauge fields (elastic fields). Outcomes of this
procedure are twofold: i) coupling of electrons with the gauge field are
generated straightforwardly, with electrons being scattered by elastic
fluctuations. In the linear regime, this interaction is identified with the
standard electron-phonon coupling. When the electron dynamics include
spin-orbit effects, we found that the elastic field couples with the spin
current, inducing novel magneto-elastic phenomena as a reflection of gauge
symmetries. We discuss several interesting scenarios that may be relevant in
the field of spintronics; ii) the dynamics of the free gauge field mimics
the weak limit of Einstein's theory of gravity, with equations that are
similar in many respects to Einstein's field equations. Hopefully, one may
get interesting insights from this analogy, considering that the theory of
elasticity can be quantized, at least in the linear regime, with phonons as
the field quanta. This topic deserves a full study by itself in future
investigations. In this paper, we will concentrate on the electronic
interactions generated by the gauge field.

The content of this paper is organized as follows: in the next Section, we
develop the general theory, imposing gauge invariance under general
space-time dependent translations. Our electronic system is described by a
simple Lagrangian density, with a free-electron structure. Local symmetry
requires the introduction of a gauge field, associated with the elastic
properties of the solid. A dominant vector phonon field is obtained in the
linear regime, for translations which are of space-like character. In
Section III, we illustrate the linear regime for the unidimensional case.
The vector phonon field is quantized yielding phonon quasi-particles, and we
identify the electron-phonon coupling comparing with standard solid state
results. In Section IV, relying on spin-orbit effects and gauge invariance
properties, we predict a novel interaction which couples the elastic field
with the electronic spin and which may be important for spin transport in a
variety of systems. Finally, in the last Section a few comments and
conclusion are added. In particular, we speculate over possible scenarios
for superconductivity, when pairing is induced by the electron-phonon
interaction, in the presence of strong magneto-elastic effects. We suggest
that our discussion is relevant for the physics of a class of iron-based
superconductors\cite{review,ironbased}.

\section{The phonon field as a gauge field}

It can be thought that electrons actually moving under the influence of a
crystalline potential live in a world whose background geometry is provided
by the crystal lattice. The background metric is the analogous of a local
Minkowskian metric in general relativity and corresponds to what the
electron effectively feels by freezing all the crystal lattice ions in their
exact equilibrium positions. Excitations, such as vibrations of the lattice
and displacements of the ions, are \ `felt ' by the electrons as
fluctuations of the background metric. This is done in close parallel with
the idea that the gravitational field can be considered as a gauge field, by
locally gauging Lorentz transformations (with local coordinate dependent
space-time translations and rotations)\cite{spingauge}. In the present case,
we want to associate a gauge field with elastic deformations in a solid. In
order to achieve this goal, we must impose invariance of the electron
Lagrangian density under coordinate dependent translations. Infinitesimal
transformation of this kind involves rigid-body displacements (constant
translations) and deformations with an associated strain field. Considering
the non-relativistic theory as our first approximation, the Lagrangian
density describing a non-relativistic Pauli electron moving in the
crystalline lattice, in units of $\hbar =1$, is given below:
\begin{equation}
\mathcal{L}=i\psi ^{\dagger }\frac{\partial \psi }{\partial t}-\frac{1}{%
2m^{\ast }}\nabla \psi ^{\dagger }\cdot \nabla \psi ~,  \label{PLD}
\end{equation}%
where $\psi $ is a Pauli spinor describing the electron and $m^{\ast }$ is
the electronic effective mass. Apart from $m^{\ast }$, the Lagrangian
density (\ref{PLD}) represents a free electron in the lattice. In the long
wavelength limit of the theory, the crystalline lattice is \ `seen ' as a
uniform mass distribution. In passing, we notice that a common viewpoint in
the description of collective excitations of a crystalline lattice, assumes
a symmetry breaking from a continuous translational symmetry to a discrete
subgroup, corresponding to translations between equivalent points in
different primitive cells. A manifest consequence of the discrete nature of
crystalline translations is the Bloch theorem, which ensures that a
translation $\mathbf{x}\rightarrow \mathbf{x}+\mathbf{a}$, being $\mathbf{a}$
an appropriate lattice translation vector, acts on the electron wavefunction
$\psi $ by means of the translation operator $\tau (\mathbf{a})=\exp [-i%
\mathbf{a}\cdot \mathbf{p}]$, where $\mathbf{p}=-i\nabla $ is the momentum
operator. The transformed wavefunction $\psi ^{\prime }=\tau (\mathbf{a}%
)\psi =\exp [-i\mathbf{a}\cdot \mathbf{p}]\psi $ differs from $\psi $ only
by a phase factor of the form $\exp [-i\mathbf{k}\cdot \mathbf{a}]$. This
concept is extended here, and we associate a gauge field to slight
distortions of the crystal lattice, inducing a space-time dependent
translation of the electronic wavefunction. Instead of a translation by a
lattice vector $\mathbf{a}$, we subject the Pauli spinor $\psi $ to an
infinitesimal translation $\delta \mathbf{a}(\mathbf{x},t)$ in space-time,
as follows
\begin{equation}
\psi ^{\prime }=\tau \lbrack \delta \mathbf{a}(\mathbf{x},t)]\psi
=[1-i\delta \mathbf{a}(\mathbf{x},t)\cdot \mathbf{p}]\Psi ~,
\label{extended bloch}
\end{equation}%
and require the invariance of the Lagrangian density (\ref{PLD}) under such
symmetry transformation. By letting the translation operator $\tau $ depend
on space-time coordinates, we raise the status of the transformation to a
local gauge symmetry, and (\ref{extended bloch}) is a generalization of the
Bloch theorem for such a case. In the non-relativistic limit, the velocity
of light is set as being infinite, and we have to introduce a velocity scale
$c_{s}$ relevant for the physics of the elastic media. This will allows us
to preserve the covariant notation of relativistic physics and to construct
a close analog of the general relativity theory, with space-time coordinates
$x^{\mu }=(c_{s}t,\mathbf{x})$. Within this modified Minkowskian metric, we
interpret $c_{s}$ as being the velocity of signals in the elastic \ `world'.
More on this later. Now that the metric is defined, we follows the gauge
prescription \cite{[ryder]}. In order to preserve the invariance of the
lagrangian density (\ref{PLD}) under the local transformations (\ref%
{extended bloch}), we must introduce gauge fields. Ordinary derivatives $%
\partial _{\mu }\equiv \partial /\partial $ $x^{\mu }$ have to be replaced
by the covariant forms $D_{\mu }$ defined as $D_{\mu }=\partial _{\mu
}+gW_{\mu }$, being $W_{\mu }$ the gauge potentials and $g$ as the coupling
constant of the theory. The gauge potentials are written in terms of the
infinitesimal generators of space-time translations, as
\begin{equation}
W_{\mu }=-iR_{\mu \nu }p^{\nu }~,  \label{defgauge}
\end{equation}%
where $p^{\nu }=i\partial ^{\nu }$ is the four-momentum operator, and $%
R_{\mu \nu }=R_{\nu \mu }$ is a symmetric second rank tensor. Einstein
summation convention over repeated indices is implied and is to be used
throughout this manuscript. Indeed, in Einstein's theory of gravity, the
potentials $R_{\mu \nu }$ are directly related to the metric tensor, at
least in the weak gravitational field regime\cite{[wheeler],[ohanian]}, but
here we identify them as the tensor components of the elasticity field. The
dynamic laws of $R_{\mu \nu }$ are similar to Einstein's field equations and
follows from the free field lagrangian density
\begin{equation}
\mathcal{L}_{GF}=-\frac{1}{4}G_{\mu \nu \beta }G^{\mu \nu \beta }=-\frac{1}{4%
}Tr\left[ G_{\mu \nu }G^{\mu \nu }\right] ,  \label{free}
\end{equation}%
with the elasticity fields $G_{\mu \nu }$ being obtained from the commutator
of the covariant derivatives \cite{[ryder]}:
\begin{eqnarray}
gG_{\mu \nu } &=&i[D_{\mu },D_{\nu }]=i(D_{\mu }D_{\nu }-D_{\nu }D_{\mu })
\notag \\
&=&ig\left( \partial _{\mu }W_{\nu }-\partial _{\nu }W_{\mu }\right) +ig^{2}
\left[ W_{\mu },W_{\nu }\right]   \notag \\
&=&g\ G_{\mu \nu \beta }\ p^{\beta }\ .  \label{defGuv}
\end{eqnarray}%
In relation (\ref{free}), the symbol $Tr$ means taking the trace over the
vector space of the field. The commutator $\left[ W_{\mu },W_{\nu }\right] $
does not vanish, and the gauge is non-abelian. Also, the theory is obviously
non-linear and invariant under the gauge transformation $R_{{\mu \nu }%
}^{\prime }=R_{\mu \nu }-\partial _{\mu }a_{\nu }-\partial _{\nu }a_{\mu }$,
being $a_{\mu }$ an arbitrary four-vector. An immediate consequence of the
definition (\ref{defGuv}) is that the fields $G_{\mu \nu }$ satisfy the
Bianchi identities, $D_{\mu }G_{\nu \lambda }+D_{\lambda }G_{\mu \nu
}+D_{\nu }G_{\lambda \mu }=0$. In the spirit of a linear theory, the
lagrangian density of the elasticity fields becomes:
\begin{equation}
\mathcal{L}_{GF}=-\frac{1}{2}[\partial _{\mu }R_{\nu \beta }\partial ^{\mu
}R^{\nu \beta }-\partial _{\mu }R_{\nu \beta }\partial ^{\nu }R^{\mu \beta
}]~.  \label{PhononLag}
\end{equation}

As a particular case, we will consider only translations in space
coordinates, but not in the time coordinate, corresponding to a
transformation $[\mathbf{x}^{\prime }=\mathbf{x}+\delta \mathbf{a}(\mathbf{x}%
,t),t^{\prime }=t]$, allowing us to write the covariant derivatives as
follows:
\begin{eqnarray}
D_{t} &\equiv &\frac{D}{Dt}=\frac{\partial }{\partial t}-igc_{s}\ {\mathbf{R}%
}\cdot \mathbf{p}~\equiv \frac{\partial }{\partial t}-ig_{0}{\mathbf{R}}%
\cdot \mathbf{p}~  \label{dt} \\
D_{i} &\equiv &\frac{D}{Dx^{i}}=\frac{\partial }{\partial x^{i}}-ig\ {%
\mathbf{R}_{i}}\cdot \mathbf{p}~,  \label{dx}
\end{eqnarray}%
where $\mathbf{R}$ and $\mathbf{R}_{i}$ are gauge vector potentials related
to the phonon and strain fields to be specified later, and $g_{0}=c_{s}g$
and $g$ are the corresponding coupling constants. The full gauge invariant
lagrangian density is then given by:
\begin{equation}
\mathcal{L}=i\psi ^{\dagger }\frac{D\psi }{Dt}-\frac{1}{2m^{\ast }}%
(D_{i}^{\dagger }\psi ^{\dagger })(D_{i}\psi )+\mathcal{L}_{GF}~.
\label{PLD1}
\end{equation}%
Note that the field $R_{\mu \nu }$ is of quadrupolar character, as in
general relativity. For space-like translations, with $g\ll g_{0}$, the
coupling with the electronic part is dominated by the vector field $\mathbf{R%
}$, which we identify with the phonon vector field $\boldsymbol{\phi }(%
\mathbf{x},t)$. This corresponds to lattice displacements in the linear
limit, with three polarization degrees of freedom. In the general case, all
components of the field are coupled and phonons will be scattered by
fluctuations of the strain field. For the strong non-linear regime, identity
of phonons may be lost. In the next section we will make our calculation
more explicit using the (1+1)-dimensional case as an illustrative example.

\section{The unidimensional example}

For sake of simplicity, suppose that space is 1-dimensional. In such a case,
the Lagrangian density of the theory takes the form:
\begin{eqnarray}
\mathcal{L} &=&\mathcal{L}_{GF}+i\psi ^{\dagger }\frac{\partial \psi }{%
\partial t}-\psi ^{\dagger }g_{0}R\frac{\partial \psi }{\partial x}-  \notag
\\
&&-\frac{1}{2m^{\ast }}[(\partial _{x}-gR_{x}\partial _{x})\psi ^{\dagger
}][(\partial _{x}-gR_{x}\partial _{x})\psi ]~.  \label{PLD2}
\end{eqnarray}%
The quantity $R$ is actually the off-diagonal component of the tensor $%
R_{\mu \nu }$, with $R=R_{01}=R_{10}$, while $R_{x}=R_{11}$. The linearized
version of the free gauge Lagrangian density now takes the form:
\begin{equation}
\mathcal{L}_{GF}=-\frac{1}{2}[(\partial _{0}R_{x})^{2}+(\partial
_{x}R)^{2}-(\partial _{0}R)^{2}-2\left( \partial _{x}R\right) \left(
\partial _{0}R_{x}\right) ]~,  \label{freephononld}
\end{equation}%
yielding the Hamiltonian density of the gauge field in a straightforward
way:
\begin{equation}
\mathcal{H}_{GF}=\frac{1}{2}\left[ \frac{1}{c_{s}^{2}}\left( \frac{\partial R%
}{\partial t}\right) ^{2}+\left( \frac{\partial R}{\partial x}\right) ^{2}-%
\frac{1}{c_{s}^{2}}\left( \frac{\partial R_{x}}{\partial t}\right) ^{2}%
\right] ~.  \label{phononH}
\end{equation}%
The fields can be decoupled if we fix the gauge with the condition:%
\begin{equation*}
\frac{\partial R_{x}}{\partial t}=0,
\end{equation*}%
showing that $R_{x}$ plays the role of a static strain field. As discussed
above, we identify the field $R$ with the phonon field $\phi $, with
longitudinal polarization for the 1-dimensional line. The Hamiltonian
density is rewritten in the form
\begin{equation*}
\mathcal{H}_{FG}=\frac{1}{2}\left[ \frac{1}{c_{s}^{2}}\left( \frac{\partial
\phi }{\partial t}\right) ^{2}+\left( \frac{\partial \phi }{\partial x}%
\right) ^{2}\right] ,
\end{equation*}%
which results in the well known linear wave equation, with wave velocity $%
c_{s}$. The velocity scale of the metric if then recognize as the velocity
of sound waves in the elastic media. Now, we want to quantize the theory,
introducing phonon variables. Following standard treatments, as the one
given in Ref. \onlinecite{[kittel]}, we expand the phonon field
(displacement operator) in terms of creation and annihilation bosonic
operators $a_{q}^{\dagger }$ and $a_{q}$ as follows:%
\begin{equation}
\phi (x,t)=\sum_{q}\sqrt{\frac{1}{2L\rho \omega _{q}}}\left[ a_{q}\
e^{i(qx-\omega _{q}t)}+a_{q}^{\dagger }\ e^{-i(qx-\omega _{q}t)}\right] ,
\label{phonon}
\end{equation}%
where $\rho $ is the mass density, $\omega _{q}$ is the energy dispersion
relation, and $L$ is the size of the system. The operator $\phi (x,t)$ in (%
\ref{phonon}) is understood to be written in the Heisenberg picture, and
since it satisfies the wave equation, it follows that $\omega _{q}$ is
linear in the wave number, $\omega _{q}=c_{s}\left\vert q\right\vert $. This
allows us to write the second quantized Hamiltonian in the form
\begin{equation*}
\hat{H}_{FG}=\int dx\mathcal{H}_{FG}=\sum_{q}~\omega _{q}a_{q}^{\dagger
}a_{q}.
\end{equation*}%
Notice that the dispersion relation $\omega _{q}=c_{s}\left\vert
q\right\vert $ here obtained, is a typical feature of long wavelength
acoustic phonons. Since the free gauge field emulates Einstein theory of
gravity, this unidimensional condensed matter system can be used to study
particular aspects of gravity and how it can be quantized. As a rewarding
point in this example, we get the standard electron-phonon interaction, once
the electron field is second-quantized. If we only consider the coupling
with the $R$ field in (\ref{PLD2}), the $\psi ^{\dagger }g_{0}R\ \partial
_{x}\psi $ term is not Hermitian. Proper symmetrization leads to%
\begin{equation}
\mathcal{L}_{I}=-\frac{1}{2}g_{0}\psi ^{\dagger }\left( \partial _{x}R\
\right) \psi \ ,  \label{symmL}
\end{equation}%
\emph{i.e.} the interaction is given in terms of the \ `dilation operator' $%
\partial _{x}R$, as shown for example in Ref. \onlinecite{[kittel]}. Next,
we expand the electron field in plane waves:%
\begin{equation*}
\psi =\sum_{k,\sigma }\ \frac{1}{\sqrt{L}}\left( e^{ikx}c_{k\sigma
}+e^{-ikx}c_{k\sigma }^{\dag }\right) \ ,
\end{equation*}%
where $\left( c_{k\sigma },c_{k\sigma }^{\dag }\right) $ are fermionic
destruction and creation operators of particles with wave number $k$ and
spin $\sigma $. Comparison with well known results in solid state theory
\cite{ziman}, allows us to identify the coupling constant of the linear
theory as%
\begin{equation*}
\frac{1}{2}g_{0}\rightarrow \frac{1}{L}U(q)\ ,
\end{equation*}%
where $U(q)=\int dx\ e^{-iqx}U\left( x\right) $ is the Fourier transform of
the lattice potential evaluated at the phonon wave number $q$. For the long
wavelength limit, $q\approx 0$, we get%
\begin{equation*}
\frac{1}{2}g_{0}=\left\langle U(x)\right\rangle \ ,
\end{equation*}%
where $\left\langle U(x)\right\rangle $ is the mean value of the lattice
potential.

We conclude this Section by remarking that we have obtained the long
wavelength phonon theory imposing gauge invariance of the electronic
lagrangian density, in which the phonon field simply appears as the gauge
field. That is, translational symmetry is of local character and not global,
and the gauge field is introduced to ensure full invariance of the theory.
In spite that we have resorted to crystalline structures to compare with
well known solid state examples, it is evident that our approach is far more
general and can be applied, in addition of crystals, to a plethora of `soft'
condensed matter systems. A contrasting point of view tells us that phonons
emerge as Goldstone bosons associated to a spontaneous breaking of the
global Galilean symmetry group. In this case, the continuous translational
symmetry reduces to the discrete translational group of the crystalline
lattice.

\section{Coupling of elastic deformations with spin-currents}

The aim of this Section is to demonstrate the existence of a coupling
between the electronic spin and the strain field of the \ crystal lattice
via the spin-orbit interaction, assuming that gauge invariance of the whole
theory applies. This coupling has been included as an \ \emph{ad hoc }
interaction in several studies of strain effects in semiconductors\cite%
{kp,qdots}, under the name of \ `spin-orbital strain effects'. The interplay
of strain, exchange, and spin-orbit coupling may also be responsible for
spin polarization in graphene, as suggested by a recent calculation\cite%
{graphene}. Below, we will show that the above coupling is of fundamental
origin. Indeed, consideration of spin-orbit interaction leads to the
electronic Hamiltonian density:%
\begin{equation}
\mathcal{H}_{SO}=-\frac{i\mu _{B}}{4m}(\nabla \psi ^{\dagger }\cdot
\boldsymbol{\sigma }\times \mathbf{E}\psi -\psi ^{\dagger }\boldsymbol{%
\sigma }\times \mathbf{E}\cdot \nabla \psi ),  \label{soc}
\end{equation}%
where $\mu _{B}$ is the Bohr magneton and $\mathbf{E}$ is the crystalline
electric field (or any external applied electric field)\cite{[dartora]}.
Such effect is named \ `spin-orbit coupling' because it has the same origin
as the analogous spin-orbit effect $-\lambda \boldsymbol{\sigma }\cdot
\mathbf{L}$ present in atoms. It is apparent that the Hamiltonian density (%
\ref{soc}) is not gauge invariant under general translations, due to the
presence of ordinary differential operators. In order to preserve the gauge
invariance of the theory, one has to replace the ordinary derivatives $%
\nabla $ by the covariant ones $D$, \ yielding:
\begin{eqnarray}
\mathcal{\tilde{H}}_{SO} &=&-\frac{i\mu _{B}}{4m}(\nabla \psi ^{\dagger
}\cdot \boldsymbol{\sigma }\times \mathbf{E}\psi -\psi ^{\dagger }%
\boldsymbol{\sigma }\times \mathbf{E}\cdot \nabla \psi )+  \notag \\
&&+i\frac{\mu _{B}}{4m}g\varepsilon _{ijk}E_{j}R_{kl}[\partial _{l}(\psi
^{\dagger })\sigma _{i}\psi -\psi ^{\dagger }\sigma _{i}\partial _{l}\psi ],
\label{soc1}
\end{eqnarray}%
where repeated indices are to be summed over and $g$ is the coupling
constant defined previously. In the extra term appearing in (\ref{soc1}),
the electronic spin is coupled to the space-like elastic field $R_{kl}$ and
to the electric field of the strained lattice. It can be written in a more
appealing form, if one defines the spin-current density as\cite%
{[dartora],[dartora2]}:%
\begin{equation*}
J_{il}=-\frac{i\mu _{B}}{2m}[(\partial _{l}\psi ^{\dagger })\sigma _{i}\psi
-\psi ^{\dagger }\sigma _{i}(\partial _{l}\psi )]~,
\end{equation*}%
thus allowing us to rewrite the additional term appearing in (\ref{soc1}) in
the following way:
\begin{equation}
\mathcal{H}_{SO}^{\prime }=-\frac{g}{2}\varepsilon _{ijk}J_{il}R_{kl}E_{j}~.
\label{soc2}
\end{equation}%
A coupling in the form $J_{il}R_{kl}$ reveals the existence of an
interaction between the spin current density $J_{il}$ and the elastic field $%
R_{kl}$. In the present contribution, we will not pursue the full
quantization of the field $R_{kl}$, that deserves a thorough study by
itself, with intrinsic difficulties related to quantize a non-linear field.
Generically, we will call it as the `strain field'. At least some immediate
physical consequences can be predicted, based on the above analysis: \ i)
electron spin-flip processes are allowed by scattering from strain field
fluctuations, as given by relation (\ref{soc2}). Thus, the thermal elastic
field may also be responsible for spin decoherence in the process of spin
transport in condensed matter; \ ii) appropriate mechanical manipulations
could be used to stimulate spin-currents in the material or to control spin
qubits in a quantum computer nanodevice, as for example the case shown in
Ref.\onlinecite{nanotube};
iii) the newly discovered spin Seebeck effect\cite{sseebeck} may also be
related to the coupling (\ref{soc2}). In a typical setup, a temperature
gradient causes a spin `voltage' across the sample, due to the generation of
a spin current. In this novel phenomenon, thermal phonons, spin-orbit
interaction and spin currents effects are interrelated. A full theoretical
explanation is still lacking, but we suggest that theoretical models based
on (\ref{soc2}) may yield some key clues in this scenario.

\section{Conclusion}

In summary, contrasting to the general understanding of phonons as being
Goldstone bosons, in this manuscript we propose that phonons are related to
a gauge field associated to local symmetry properties (gauge bosons). The
application of Goldstone theorem is subdued by the gauge invariance
principle, which provides us with a more general approach. In fact, the
Goldstone boson is associated with the spontaneous breaking of a global
symmetry, while gauge theories deal with local symmetry properties.

This local symmetry principle is applied to a condensed matter system
consisting of electrons moving in a solid with elastic properties. We adopt
for electrons a minimal model, that only includes the kinetic energy and the
effective mass. In spite that infinitesimal generators of translations
commute, the gauge field is not abelian, due to the space-time dependent
character of generalized translations. Electronic interactions are generated
through couplings with the gauge field, when one requires full gauge
invariance. The background medium is rigged with a space-time structure that
mimics the Minkowskian metric, with the sound velocity playing the role of
the velocity of light. The gauge field describes elastic properties of the
solid and supports collective excitations in the form of elastic waves. The
free field is in principle non-linear, and emulates Einstein's theory of
General Relativity. When we consider space-like translations in the linear
limit of the theory, a vector branch of the field decouples from the other
components, forming the \ `phonon' field. We elaborate this case in \emph{%
1-dim}, showing that the theory reduces to the long wavelength limit of
acoustic phonons, with a dispersion relation of the form $\omega _{k}=c_{s}k$%
, with $c_{s}$ being the sound velocity. The resulting electron-phonon
interaction is full identified, comparing with standard treatments in solid
state theory. Most important, novel predictions are obtained when we enforce
gauge invariance on the spin-orbit interaction, revealing the existence of a
coupling between the spin current density and the elastic field. The
interplay of elastic properties and spin transport is certainly a key
question in the field of spintronics.

In closing this section, we comment on a possible pairing scenario for
superconductivity, when the superconducting state is mediated by the
electron-phonon interaction. If the spin-orbit coupling is not negligible,
an interaction term such as (\ref{soc2}) will introduce spin flip in the
pairing mechanism, since phonons scattered by strain field fluctuations will
couple with the spin current. As a result, Cooper pairs will acquire singlet
and triplet admixtures. This will imply a mixed symmetry of the order
parameter. Furthermore, coupling between elastic and magnetic fluctuations
may open a road for enhancing the electron-phonon interaction in materials
where the magnetoelastic effect plays an important role. We suggest that
this discussion is relevant for the physics of the recently discovered
iron-based superconductors\cite{review,ironbased}. With unusual high
transition temperatures, they exhibit a superconducting phase that is very
close to structural and magnetic transitions, giving a strong hint that the
above phenomena are interrelated\cite{magnetoelastic}. Our theory proposes
that the above connection is a fundamental trait that comes from general
principles.\newline


\textbf{Acknowledgements}

One of the authors (C.A.D.) would like to thank the Brazilian agency CNPq
for partial financial support.

\end{document}